\documentclass[12pt]{article}
\usepackage{amsmath,amsfonts,epsf}
\usepackage{graphicx}
\usepackage{grffile}
\input epsf

\makeatletter\@addtoreset{equation}{section}\makeatother

\def\be{\begin{equation}}
\def\ee{\end{equation}}
\def\bea{\begin{eqnarray}}
\def\eea{\end{eqnarray}}
\def\ie{\begin{equation}\begin{aligned}}
\def\fe{\end{aligned}\end{equation}}

\newcommand{\A}{{\alpha}}
\newcommand{\B}{{\beta}}

\newcommand{\da}{{\dot\alpha}}
\newcommand{\db}{{\dot\beta}}

\makeatletter\@addtoreset{equation}{section}\makeatother

\renewcommand{\title}[1]{\vbox{\center\LARGE{#1}}\vspace{5mm}}
\renewcommand{\author}[1]{\vbox{\center#1}\vspace{5mm}}
\newcommand{\address}[1]{\vbox{\center\em#1}}
\newcommand{\email}[1]{\vbox{\center\tt#1}\vspace{5mm}}

\begin{document}
\begin{titlepage}
\begin{center}
\hfill \\
\hfill \\
\vskip 1cm

\title{Higher Spins in AdS and Twistorial Holography}

\author{Simone Giombi$^{1,a}$ and
Xi Yin$^{1,2,b}$}

\address{${}^1$Center for the Fundamental Laws of Nature
\\
Jefferson Physical Laboratory, Harvard University,\\
Cambridge, MA 02138 USA}
\address{
${}^2$School of Natural Sciences,
Institute for Advanced Study,\\
Princeton, NJ 08540 USA}

\email{$^a$giombi@physics.harvard.edu,
$^b$xiyin@fas.harvard.edu}

\end{center}

\abstract{ In this paper we simplify and extend previous work on three-point functions
in Vasiliev's higher spin gauge theory in $AdS_4$. We work
in a gauge in which the space-time dependence of Vasiliev's master fields is gauged away completely, leaving only the internal twistor-like variables. The correlation functions of boundary operators can be easily computed in this gauge. We find complete agreement of the tree level three point functions of higher spin currents in Vasiliev's theory with the conjectured dual free $O(N)$ vector theory.
}

\vfill

\end{titlepage}


\section{Introduction}

The conjectured duality between Vasiliev's minimal bosonic higher spin gauge theory in $AdS_4$ and the free/critical $O(N)$ vector model \cite{Klebanov:2002ja} (for earlier closely related work, see \cite{Sezgin:2002rt}) is an example of AdS/CFT duality \cite{Maldacena:1997re, Gubser:1998bc, Witten:1998qj} which is remarkable for a number of reasons. Firstly, the bulk higher spin gauge theory is analogous to the tensionless limit of string field theories in AdS space, but has explicitly known classical equations of motion. Secondly, the conjecture provides the first explicit holographic dual of a free (gauge) theory. Thirdly, the conjecture provides the first precise holographic dual of a CFT that can be realized in the real world, namely the critical $O(N)$ vector model (for small values of $N$).
In a recent work by the authors \cite{Giombi:2009wh}, concrete evidence in support of this conjecture was found by computing tree level three point functions of currents from the bulk theory, specialized to the case where one of the currents is a scalar operator, and comparing to the boundary CFT, for both $\Delta=1$ and $\Delta=2$ boundary conditions. However, the method of computation in \cite{Giombi:2009wh} was laborious and difficult to extend to more complicated correlation functions. It was also difficult to recognize the rather simple structures of the boundary CFT in the messy details of the bulk computation.

In this paper, we will compute the holographic correlation functions in a different gauge \cite{Vasiliev:1990bu}  (see also \cite{Bolotin:1999fa, Bekaert:2005vh, Sezgin:2005pv}), in which the spacetime dependence of Vasiliev's master fields are eliminated completely, and one only needs to work with the internal twistor-like variables. We will refer to it as the ``$W=0$ gauge". We will find drastic simplification in the computation of three point functions. In fact, one no longer needs to explicitly perform the integration over the bulk $AdS_4$, which is entirely encoded in the star product of the master fields in the new gauge. The boundary-to-bulk propagators for the higher spin fields are essentially given by delta functions on the twistor space, and the resulting correlation function is represented as a contour integral on the twistor variables. We will find a completely explicit answer for the three point functions of all higher spin currents, which precisely agrees with that of the free $O(N)$ vector theory.

We would like to emphasize that this agreement is a highly nontrivial test of the structure of higher derivative couplings of Vasiliev theory. For instance, the three point function of the stress energy tensor $\langle TTT\rangle$ of a three dimensional CFT is constrained by conformal symmetry up to a linear combination of two tensor structures \cite{Osborn:1993cr}, corresponding to that of a free massless scalar field and a free fermion, respectively. In particular, the tree level contribution to $\langle TTT\rangle$ from Einstein-Hilbert Lagrangian in the bulk is a linear combination of both \cite{Arutyunov:1999nw}. A holographic dual of free scalars therefore must involve higher derivative couplings in the graviton sector. Our result confirms that Vasiliev theory has precisely the higher derivative couplings to produce the correct three point functions.

\section{The $W=0$ gauge}

Vasiliev's minimal bosonic higher spin gauge theory in $AdS_4$ \cite{Vasiliev:1992av,Vasiliev:1995dn, Vasiliev:1999ba} is formulated in terms of the master fields $W=W_\mu dx^\mu$, $S=S_\A dz^\A + S_\da d\bar z^\da$ and $B$, which are functions of the spacetime coordinates $x^\mu$ and the internal variables $(Y,Z)=(y_\A, \bar y_\da, z_\A, \bar z_\da)$.\footnote{Sometimes we will use the notation $\hat z_\A$ instead of $z_\A$ for the internal variables to avoid confusion with the Poincar\'e radial coordinate $z$.}
The classical equation of motion takes the form
\ie\label{cleom}
& d_xW+W*W=0,\\
& d_Z W + d_x S + W*S+S*W=0,\\
& d_ZS+S*S = B*(K dz^2 + \bar K d\bar z^2),\\
& d_x B + W*B-B*\bar\pi(W)=0,\\
& d_Z B + S*B - B*\bar\pi(S)=0.
\fe
Here $K=e^{z^\A y_\A}$ and $\bar K=e^{\bar z^\da \bar y_\da}$ are Kleinians, and we define $dz^2=\frac{1}{2}dz^{\alpha} dz_{\alpha}$, $d\bar z^2=\frac{1}{2}d\bar z^{\dot \alpha} d\bar z_{\dot \alpha}$; $\bar\pi$ is the operation $(y,\bar y,z,\bar z,dz,d\bar z)\mapsto (y,-\bar y,z,-\bar z,dz,-d\bar z)$.
We shall refer the reader to \cite{Giombi:2009wh} and references therein for a review of Vasiliev's theory and the detailed conventions. Throughout most of this paper we will be working with the type A model of \cite{Sezgin:2003pt}, where the bulk scalar is chosen to be parity even, while commenting on the type B model (with a parity odd scalar) briefly towards the end. The minimal bosonic type A model can be defined by projecting the fields onto the components invariant under the symmetry
\ie\label{proja}
&W(x|y,\bar y,z,\bar z) \to -W(x|iy,i\bar y,-iz,-i\bar z),\\
&S(x|y,\bar y,z,\bar z,dz,d\bar z) \to -S(x|iy,i\bar y,-iz,-i\bar z,-idz,-id\bar z),\\
&B(x|y,\bar y,z,\bar z) \to B(x|iy,-i\bar y,-iz,i\bar z),\\
\fe
and consequently only the even integer spin fields are retained.

Because $W$ is a flat connection in spacetime, at least locally we can always go to a gauge in which $W$ is set to zero. We will denote by $S'$ and $B'$ the corresponding master fields in this gauge. The equations of motion then states that $S'$ and $B'$ are independent of the spacetime coordinates $x^\mu$, and are functions of $Y,Z$ only. Explicitly, we can write
\ie
&W(x|Y,Z) = g^{-1}(x|Y,Z)*d_xg(x|Y,Z),\\
&S(x|Y,Z) = g^{-1}(x|Y,Z)*d_Zg(x|Y,Z)+ g^{-1}(x|Y,Z)*S'(Y,Z)*g(x|Y,Z),\\
&B(x|Y,Z) = g^{-1}(x|Y,Z)*B'(Y,Z)*\pi(g(x|Y,Z)).
\fe
Here $g^{-1}$ stands for the $*$-inverse of $g$. The equations for $S'$ and $B'$
\ie
& d_ZS'+S'*S' = B'*(K dz^2 + \bar K d\bar z^2),\\
& d_Z B'+ S'*B' - B'*\pi(S')=0,
\fe
are now much simpler to solve. In order to extract holographic correlation functions, however, we must go back to the standard ``physical" gauge in the end, and extract boundary expectation of the fields.

As in \cite{Giombi:2009wh}, our strategy of computing the $n$-point correlation functions is to take $n-1$ higher spin currents (inserted at $n-1$ points $\vec x_1, \vec x_2, \cdots, \vec x_{n-1}$ on the boundary) as sources for the bulk fields, and solve for the $(n-1)$-th order field in the bulk by sewing together the boundary-to-bulk propagators using the equation of motion. The tree-level correlation function of the higher spin currents will then be extracted from the expectation value of $B(x|y_\A,\bar y_\da=z_\A=\bar z_\da=0)$ near the boundary, say at a point $\vec x_n$.\footnote{Note that $B$ contains the generalized Weyl curvatures of the higher spin gauge fields. Nevertheless, to leading order in the Poincar\'e radial coordinate $z$ near the boundary $z=0$, the generalized Weyl curvature is proportional to the gauge field itself (in the gauge of \cite{Giombi:2009wh}, see in particular eq.~(3.60) of \cite{Giombi:2009wh}), and the correlation function of the current can be directly extracted from $B(x|y_\A,\bar y_\da=z_\A=\bar z_\da=0)$.}

Working in perturbation theory, we start by writing the $AdS_4$ vacuum solution as
\ie
W_0(x|y,\bar y) = L^{-1}(x|y,\bar y)*dL(x|y,\bar y),
\fe
for a gauge function $L(x|Y)$.
One begins with the linearized field $B^{(1)}(x,Y)$, and transform it to the $W=0$ gauge $B'^{(1)}(Y)$. We can then solve the linearized field $S'^{(1)}$ from
\ie
d_Z S'^{(1)}= B'^{(1)}*(Kdz^2+\bar Kd\bar z^2).
\fe
Explicitly, the solution is
\ie
S'^{(1)} &=-z_\A dz^\A \int_0^1 dt\,t (B'^{(1)}*K)|_{z\to tz} + c.c.\\
&= -z_\A dz^\A \int_0^1 dt\,t B'^{(1)}(-tz,\bar y)K(t) + c.c.
\fe
Here we have made the gauge choice $S'|_{Z=0}=0$, following \cite{Vasiliev:1995dn, Vasiliev:1999ba}.
Next, the quadratic order fields $B'^{(2)}$ and $S'^{(2)}$ can be solved from
\ie
& d_ZB'^{(2)} = -S'^{(1)}*B'^{(1)}+B'^{(1)}*\pi(S'^{(1)}),\\
& d_Z S'^{(2)} = -S'^{(1)}*S'^{(1)} + B'^{(2)}*(Kdz^2+\bar K d\bar z^2).
\fe
For tree-level three point functions it suffices to solve for $B'^{(2)}(Y,Z)$ only, which is explicitly given in terms of the linearized fields by
\ie
&B'^{(2)}(y,\bar y, z,\bar z) = -z^\A \int_0^1 dt \left[ S_\A'^{(1)}*B'^{(1)} - B'^{(1)}*\bar\pi(S_\A'^{(1)}) \right]_{z\to tz} + c.c.
\\
&= z^\A \int_0^1 dt \left[ \left(z_\A \int_0^1 d\eta \,\eta B'^{(1)}(-\eta z,\bar y)K(\eta) \right)*B'^{(1)}(y,\bar y) \right.
\\
&~~~~\left.- B'^{(1)}(y,\bar y)*\left(z_\A \int_0^1 d\eta \,\eta B'^{(1)}(\eta z,\bar y)K(\eta) \right) \right]_{z\to tz} + c.c.
\\
&=  z^\A \int_0^1 dt\int_0^1 d\eta \,\eta  \left[ B'^{(1)}(-\eta z,\bar y)K(\eta) * \partial_\A B'^{(1)}(y,\bar y) - \partial_\A B'^{(1)}(y,\bar y)* B'^{(1)}(\eta z,\bar y)K(\eta) \right]_{z\to tz} + c.c.
\\
&=  z^\A \int_0^1 dt\int_0^1 d\eta \,\eta \int du dv e^{uv+\bar u\bar v}\partial_\A B'^{(1)}(y+v,\bar y+\bar v)
\\
&~~~\times  \left[ B'^{(1)}(-\eta (tz+u),\bar y+\bar u)e^{\eta (tz+u)(y+ u)}  - B'^{(1)}(\eta (tz+u),\bar y-\bar u)e^{\eta (tz+u)(y-u)} \right]+ c.c.
\\
&= -2 z^\A \int_0^1 dt\int_0^1 d\eta \,\eta \int d^4u d^4v \partial_\A B'^{(1)}(v,\bar v)B'^{(1)}(-\eta u,\bar u)
e^{(u-tz)(v-y)+ \bar u(\bar v-\bar y)+\eta u y} \sinh\left( \bar y\bar v+\eta t uz \right)
+c.c.
\\
&= -2 z^\A \int_0^1 dt\int_0^1 d\eta \,\eta^{-1} \int d^4u d^4v B'^{(1)}(u,\bar u)\partial_\A B'^{(1)}(v,\bar v)
e^{({u\over\eta}+tz)(y-v)+ \bar u(\bar v-\bar y)- u y} \sinh\left( \bar y\bar v- t uz \right)
+c.c.
\\
&= 2 \int_0^1 dt\int_0^\infty d\eta \int d^4u d^4v e^{-uv+\bar u\bar v} B'^{(1)}(u,\bar u) B'^{(1)}(v,\bar v) (zu)
e^{(\eta u+tz)(y-v) + \bar y\bar u} \sinh\left( \bar y\bar v- t uz \right)
+c.c.
\\
& = 2 \int d^4u d^4v e^{-uv+\bar u\bar v} B'^{(1)}(u,\bar u) B'^{(1)}(v,\bar v)
f(y,\bar y,z;U,V)
+c.c.
\label{Bprime2}
\fe
In the steps above we have made several redefinitions on the variables $u,\bar u,v,\bar v$ and $\eta$. In the last step we defined the function
\ie
f(y,\bar y,z;U,V) = \int_0^1 dt\int_0^\infty d\eta (zu)
e^{(\eta u+tz)(y-v) + \bar y\bar u} \sinh\left( \bar y\bar v+ t zu \right).
\fe
Finally, we will be able to recover the second order $B$ field in the standard ``physical" gauge by
\ie\label{btwo}
B^{(2)}(x|Y,Z)& = L^{-1}(x,Y) * B'^{(2)}(Y,Z) * \pi(L(x,Y))\\
&~~~- \epsilon^{(1)}(x,Y,Z)*B^{(1)}(x,Y) + B^{(1)}(x,Y)*\pi(\epsilon^{(1)}(x,Y,Z)),
\fe
and then take $x= (\vec x,z\to 0)$ while restricting to $\bar y = \hat z=\hat {\bar z}=0$ to extract the three point function. Here $\epsilon^{(1)}(x,Y,Z)$ is a first order correction to the gauge function $L(x,Y)$. To understand its effect, let us consider the linearized fields
\ie
&W^{(1)}(x|Y,Z) = D_0 \epsilon^{(1)}(x,Y,Z) ,\\
&S^{(1)}(x|Y,Z) = L^{-1}(x,Y)*S'^{(1)}(Y,Z)*L(x,Y) + d_Z \epsilon^{(1)}(x|Y,Z).
\fe
Near the boundary $z\to 0$, the spin-$s$ component of $\Omega(\vec x,z|Y)$ falls off like $z^s$, whereas
the spin-$s$ component of $B(\vec x,z|Y)$ falls off like $z^{s+1}$. It is then natural to impose the $z^s$ fall-off condition on the spin-$s$ component of the gauge function $\epsilon^{(1)}(x|Y,Z)$. So generically we expect the ``gauge correction" in (\ref{btwo}) to fall off like $z^{s_1+s_2+1}$, which does not affect the leading boundary behavior of the spin-$s'$ component of $B^{(2)}$, if $s'<s_1+s_2$. Given three spins $s_1, s_2, s_3$ (not all zero), we can always choose two sources, say $s_1, s_2$, so that $s_3<s_1+s_2$. In this case, we can drop the linear gauge function in (\ref{btwo}), for the purpose of extracting the boundary correlation function.

Note that in going back to the physical gauge, $\epsilon^{(1)}(x|Y,Z)$ should be chosen so that the gauge condition $S|_{Z=0}=0$ is preserved. There are additional gauge ambiguities of the form $\tilde\epsilon^{(1)}(x|Y)$, under which $\Omega(x|Y)$ transforms by $\delta \Omega^{(1)}(x|Y) = D_0\tilde\epsilon^{(1)}(x|Y)$. For the purpose of extracting three point functions from the boundary expectation value, it suffices to consider the second order $B$-field, restricted to $\bar y_\da=z_\A=\bar z_\da=0$ (which contains the self-dual part of the higher spin Weyl curvature tensor). Its gauge variation under $\tilde\epsilon^{(1)}(x|Y)$ is given by
\ie\label{bvar}
\delta B^{(2)}(x|y,\bar y=z=\bar z=0) = -\tilde \epsilon^{(1)}(x|Y) * B^{(1)}(x|Y) + B^{(1)}(x|Y)*\pi(\tilde\epsilon^{(1)}(x|Y))
\fe
The spin $s_1$-components of $\tilde\epsilon^{(1)}$ consists of terms of the form $\tilde\epsilon^{(s_1-1+k,s_1-1-k)}$, $1-s_1\leq k\leq s_1-1$, where the superscripts indicate the degrees in $y$ and $\bar y$ respectively. The spin $s_2$-components of $B^{(1)}$ consists of terms of the form $B^{(2s_2+n,n)}$ and $B^{(n,2s_2+n)}$, $n\geq 0$. It is then easily seen that after contracting all the $\bar y$'s under the $*$ product on the RHS of (\ref{bvar}), $\delta B^{(2)}(x|y,\bar y=z=\bar z=0)$ may be nonzero only for components of spin $s_3<s_1+s_2$ (i.e. terms of degree $2s_3$ in $y$). We have argued previously that the falloff behavior of the gauge functions near the boundary of $AdS$ is such that the leading boundary behavior of $B^{(2)}$ is not affected when $s_3<s_1+s_2$. Therefore, there is no ambiguity due to $\tilde\epsilon^{(1)}$ in extracting the boundary correlators for all spins.

\section{The gauge function and boundary-to-bulk propagator}

To carry out the computation in $W=0$ gauge explicitly, first we shall write down the gauge function
\ie
L(x,Y) = {\bf P} \exp_*\left(-\int^{x_0}_x W_0^{\mu}(x'|Y) dx'_\mu\right)
\fe
where the $*$-exponential is path ordered, from $x=(\vec x,z)$ to a base point $x_0=(\vec x_0,z_0)$. The $AdS_4$ vacuum solution is given by $W_0=e_0+\omega_0^L$, where $e_0$ and $\omega_0^L$ are the vielbein and spin connection of $AdS_4$, which in our conventions \cite{Giombi:2009wh} take the form (in Poincar\'e coordinates)
\ie
&\omega_0^L = {1\over 8} {dx^i\over z} \left[ (\sigma^{iz})_{\A\B}y^\A y^\B
+(\sigma^{iz})_{\da\db}\bar y^\da \bar y^\db \right],\\
& e_0 = {1\over 4} {dx_\mu\over z} \sigma^\mu_{\A\db}y^\A \bar y^\db.
\fe
If we choose the straight contour $x(t) = (1-t)x_0+t x$, then the value of $W_0$ along different points on the contour
$*$-commute with one another, and we can write simply
\ie
L(x,Y) &= \exp_* \left[ (x-x_0)_\mu \int_0^1 dt W_0^\mu((1-t)x_0+t x|y,\bar y) \right]
\\
&= \exp_* \left[ \int_0^1 dt {(x-x_0)_\mu\over (1-t) z_0 + t z} \bar\omega_0^\mu(y,\bar y) \right]
\\
&= \exp_* \left[ -{1\over 8}\int_0^1 {dt\over (1-t) z_0 + t z} \left( y({\bf x}-{\bf x_0})\sigma^z y +
\bar y({\bf x}-{\bf x_0})\sigma^z \bar y + 2y ({\bf x}-{\bf x}_0) \bar y \right) \right]
\\
&= \exp_* \left[ -{1\over 8} \ln(z/z_0)\left( (y-\bar y\sigma^z){{\bf x}-{\bf x_0}\over z-z_0}(\sigma^z y+\bar y) +
2 y\sigma^z\bar y \right) \right].
\fe
Here we have 
introduced the notation ${\bf x}=x^{\mu} \sigma_{\mu}=x^i \sigma_i + z \sigma^z$.

Generally, given a symmetric matrix $M$, one can calculate the $*$-exponential
\ie
\exp_*\left( {t\over 2} Y M Y \right) = \exp\left[{1\over 2}Y \Omega(t) Y + f(t)\right]
\fe
where the symmetric matrix $\Omega(t)$ and function $f(t)$ satisfy
\ie
&{d\Omega(t)\over dt} = (1-\Omega(t)) M (1+\Omega(t)),\\
&{df(t)\over dt} = -{1\over 2}{\rm Tr}(M\Omega(t)).
\fe
The solution is
\ie
\Omega(t) = \tanh(tM),~~~~f(t) =-{1\over 2}{\rm Tr} \ln\cosh(tM).
\fe
So the result for the $*$-exponential is
\ie
\exp_*({1\over 2}YMY) = \left[\det(\cosh M)\right]^{-{1\over 2}} \exp\left[{1\over 2}Y(\tanh M) Y\right].
\fe
Applying this formula to $L(x,Y)$, working in the basis $(y,\bar y)$, we can write $M$ as
\ie
M(x) = -{1\over 4}{\ln(z/z_0)\over z-z_0} \left(\begin{matrix} (\vec{x}\cdot \vec{\sigma}-\vec{x}_0\cdot \vec{\sigma})\sigma^z & {\bf x}-{\bf x_0}\\
{\bf x}-{\bf x_0} & (\vec{x}\cdot \vec{\sigma}-\vec{x}_0\cdot \vec{\sigma})\sigma^z \end{matrix} \right)
\fe
It is convenient to choose the base point to be $\vec x_0 =0,z_0=1$, so that
\ie
M(x) = -{\ln z\over 4(z-1)} \left(\begin{matrix} \vec{x}\cdot \vec{\sigma}\sigma^z & \vec{x}\cdot \vec{\sigma}+(z-1)\sigma^z\\
\vec{x}\cdot \vec{\sigma}+(z-1)\sigma^z &\vec{x}\cdot \vec{\sigma}\sigma^z \end{matrix} \right)
\fe
and then
\ie
&L(x,Y) =\left[\det(\cosh M)\right]^{-{1\over 2}} \exp\left[{1\over 2}Y(\tanh M) Y\right],\\
&L^{-1}(x,Y) = \left[\det(\cosh M)\right]^{-{1\over 2}} \exp\left[-{1\over 2}Y(\tanh M) Y\right].
\fe
Our goal will be to extract the correlation function from the expectation value of a bulk field near a boundary point, given a number of boundary sources. By translation invariance we can choose the boundary point to be at $\vec x=0$, near which the bulk field will be evaluated. In other words, we are choosing the $\vec x$ Poincar\'e coordinate of the boundary point to coincide with that of the base point in the definition of $L(x,Y)$. At $\vec x=0$ and nonzero values of $z$, we have
\ie
&M = - {\ln z\over 4} \left(\begin{matrix} 0 & \sigma^z\\
\sigma^z & 0 \end{matrix} \right),\\
&\cosh M = \cosh(\ln z/4) = {z^{1\over 4}+z^{-{1\over 4}}\over 2}{\bf 1},\\
&\tanh M = {1-z^{1\over 2}\over 1+z^{1\over 2}} \left(\begin{matrix} 0 & \sigma^z\\
\sigma^z & 0 \end{matrix} \right).
\fe
So
\ie
& L^{\pm 1}(\vec x=0,z, Y) = {4\over z^{-{1\over 2}}+2+z^{1\over 2}} \exp \left( \pm {1-z^{1/2}\over 1+z^{1/2}} y\sigma^z \bar y \right).
\fe

By definition, $L(x_0,Y)=1$, at the base point $x_0^\mu=(\vec x_0,z)$,. So the linearized field in the $W=0$ gauge is simply
\ie
B'^{(1)}(Y) = B^{(1)}(x_0|Y).
\fe
Explicitly, using the formulae derived in \cite{Giombi:2009wh} (see eq.~(3.31) and eq.~(3.33) of \cite{Giombi:2009wh}), the boundary-to-bulk propagator for the spin-$s$ component of $B'$ corresponding to a boundary source located at $\vec x =0$ with a null polarization vector $\vec\varepsilon$ is given by
\ie
B'^{(1)}_{(s)}(y,\bar y) = {(y(\vec{x}_0 \cdot \vec{\sigma}+\sigma^z ){\slash\!\!\!\varepsilon}\sigma^z(\vec{x}_0 \cdot \vec{\sigma}+\sigma^z) y)^s
\over (x_0^2+1)^{2s+1}}e^{-y(\sigma^z-2{\vec{x}_0 \cdot \vec{\sigma}+\sigma^z\over x_0^2+1})\bar y}+c.c.
\fe
Alternatively, if we fix the base point to be at $(0,z=1)$ and the source at $\vec x_0$, then in the $W=0$ gauge we have
\ie
B'^{(1)}_{(s)}(y,\bar y) = {(y(\vec{x}_0 \cdot \vec{\sigma}-\sigma^z ){\slash\!\!\!\varepsilon}\sigma^z(\vec{x}_0 \cdot \vec{\sigma}-\sigma^z) y)^s
\over (x_0^2+1)^{2s+1}}e^{-y(\sigma^z+2{\vec{x}_0 \cdot \vec{\sigma}-\sigma^z\over x_0^2+1})\bar y}+c.c.
\fe
It will be useful to express the null polarization vector as a spinor bilinear $({\slash\!\!\!\varepsilon}\sigma^z)_{\da\db} =\bar \lambda_\da \bar\lambda_\db$. In our conventions, we can also write $\bar\lambda = \sigma^z \lambda$. We can then construct a generating function for the boundary-to-bulk propagator associated with currents of all spins as
\ie
B'^{(1)}(y,\bar y) &= {1\over x_0^2+1} \exp\left[ -y(\sigma^z+2{\vec{x}_0 \cdot \vec{\sigma}-\sigma^z\over x_0^2+1})\bar y
- 2\bar\lambda {\vec{x}_0 \cdot \vec{\sigma}-\sigma^z\over x_0^2+1} y \right] + c.c.
\\
&= {1\over x_0^2+1} \exp\left[ -y\sigma^z \bar y-2y{\vec{x}_0 \cdot \vec{\sigma}-\sigma^z\over x_0^2+1}(\bar y+\bar\lambda) \right] + c.c.
\fe
Keep in mind that we should in fact only select the part of this generating function which is even in $\lambda$, because the theory describe all the {\it integer} spins.\footnote{Equivalently, recall that consistency of the purely bosonic Vasiliev's equations require the constraints $W(x|-Y,-Z)=W(x|Y,Z)$, $S_{\alpha}(x|-Y,-Z)=-S_{\alpha}(x|Y,Z)$, $S_{\dot\alpha}(x|-Y,-Z)=-S_{\dot\alpha}(x|Y,Z)$ and $B(x|Y,Z)=B(x|-Y,-Z)$.}

Once we solve the second order field in the $W=0$ gauge, the expectation value of $B$ in the standard gauge at $\vec x=0$, near $z=0$, is recovered from
\ie
&B^{(2)}(\vec x =0,z\to 0,y,\bar y,z,\bar z) = L^{-1}(\vec x =0,z\to 0,y,\bar y)*B'^{(2)}(y,\bar y,z,\bar z)*L(\vec x =0,z\to 0,y,-\bar y)
\label{B2phys}
\fe
Given a function $f(Y,Z)$, let us consider the twisted adjoint action by $L$, evaluated near the boundary of $AdS_4$,
\ie
&F(z,Y,Z)=L^{-1}(\vec x=0,z\to 0, Y)*f(Y,Z)*\pi(L(\vec x=0,z\to 0, Y))\\
&\simeq  16 z \exp \left(- {1-z^{1/2}\over 1+z^{1/2}} y\sigma^z \bar y \right)*
f(y,\bar y,\hat z,\bar z)
*\exp \left(- {1-z^{1/2}\over 1+z^{1/2}} y\sigma^z \bar y \right)
\\
&=16z \int d^4 u d^4 v d^4u' d^4v' \exp\left(uv+\bar u\bar v+u'v'+\bar u'\bar v'\right) \exp \left(- {1-z^{1/2}\over 1+z^{1/2}} (y+u+u')\sigma^z (\bar y+\bar u+\bar u') \right)\\
&~~~\times
f(y+v+u',\bar y+\bar v+\bar u',\hat z-v+u',\bar z-\bar v+\bar u')
\exp \left(- {1-z^{1/2}\over 1+z^{1/2}} (y+v')\sigma^z (\bar y+\bar v') \right)
\fe
In the second line we have dropped subleading terms in $z$ in the overall factor, which do not affect the boundary expectation value of fields of various spins. For the purpose of extracting the three-point functions of the currents, we may restrict to $\bar y=\hat z=\bar z=0$ while keeping the dependence on $y$ only. Denote by $F(z,y_\A)\equiv F(z,y_\A,\bar y_\da=z_\A=\bar z_\da=0)$, we have
\ie
&F(z,y_\A)
=16z \int d^4 u d^4 v d^4u' d^4v' \exp\left(u(v-{y\over2})+\bar u\bar v+(u'-{y\over2})v'+\bar u'\bar v'\right) \\
&~~~\times\exp \left(- {1-z^{1/2}\over 1+z^{1/2}} ({y\over2}+u+u')\sigma^z (\bar u+\bar u') \right)
\exp \left(- {1-z^{1/2}\over 1+z^{1/2}} (y+v')\sigma^z \bar v' \right)\\
&~~~\times
f(v+u',\bar v+\bar u',-v+u',-\bar v+\bar u')
\\
&=16z \int d^4 u d^4 v d^4u' d^4v' \exp\left[{y\over 2}\left(u-u'-v' - {1-z^{1/2}\over 1+z^{1/2}} \sigma^z(\bar u+2\bar v') \right) - {1-z^{1/2}\over 1+z^{1/2}} (u\sigma^z \bar u +v'\sigma^z \bar v' ) \right]\\
&~~~\times \exp\left(uv+\bar u\bar v+u'v'+\bar u'\bar v' - u' v-\bar u' \bar v\right)
f(v+u',\bar v+\bar u',-v+u',-\bar v+\bar u')
\\
&=z \int d^4 u d^4v d^4 p d^4q \exp\left[{y\over 2}\left(u-{p+q\over 2}-v - {1-z^{1/2}\over 1+z^{1/2}} \sigma^z(\bar u+2\bar v) \right) - {1-z^{1/2}\over 1+z^{1/2}} (u\sigma^z \bar u +v\sigma^z \bar v ) \right]\\
&~~~\times \exp\left({u(p-q)+\bar u(\bar p-\bar q)+(p+q)v+(\bar p+\bar q)\bar v +pq+\bar p\bar q\over 2}\right)
f(p,\bar p,q,\bar q).
\fe
The functions $f(Y,Z)$ that shows up in the computation of $B^{(2)}$ depend either only on $z_\A$ or only on $\bar z_\da$ (see eq. (\ref{Bprime2})). We will treat the two cases separately. First, consider the case where $f(y,\bar y,z,\bar z)=f(y,\bar y,z)$ is independent of $\bar z_\da$. Then
\ie
&F(z,y_\A)
=4 z \int d^4 u d^2v d^4 p d^2q \exp\left[{y\over 2}\left(u-{p+q\over 2}-v - {1-z^{1/2}\over 1+z^{1/2}} \sigma^z(2\bar p-\bar u) \right) \right.\\
&~~~\left.- {1-z^{1/2}\over 1+z^{1/2}} ((u-v)\sigma^z \bar u +v\sigma^z \bar p ) \right] \exp\left({(u-v)p-(u+v)q+2\bar u \bar p +pq\over 2}\right)
f(p,\bar p,q)
\\
&=z \int d^2 p d^2q \exp\left\{ \left[ 1+ \left(1+z^{1\over 2}\over 1-z^{1\over 2}\right)^2 \right] {pq\over 2} + \left[ 1- \left(1+z^{1\over2}\over 1-z^{1\over2}\right)^2 \right] {yq\over 2} \right\}
f\left(p,-{1+z^{1\over2}\over 1-z^{1\over2}}\sigma^z q,q\right)
\\
&\to z \int d^2 p d^2q \,e^{  (1+\epsilon)pq  - 2\sqrt{z} yq }
f\left(p,-\sigma^z q,q\right)
\fe
In the last step, we have taken the limit $z\to 0$ while keeping $\sqrt{z} y$ fixed. $\epsilon\sim \sqrt{z}$ is understood as a small positive number that will be taken to zero at the end. For the moment, we need to keep it nonzero to regularize integrals that appear in $*$-products.

Now consider the other case, where $f(y,\bar y,z,\bar z) = \bar f(y,\bar y,\bar z)$ is independent of $z_\A$. Then
\ie
&F(z,y_\A) = 4z \int d^4 u d^2\bar v d^4 p d^2\bar q \exp\left[{y\over 2}\left(2u-p - {1-z^{1/2}\over 1+z^{1/2}} \sigma^z(\bar u+\bar v) \right) - {1-z^{1/2}\over 1+z^{1/2}} (u\sigma^z (\bar u-\bar v) +p\sigma^z \bar v ) \right]\\
&~~~\times \exp\left({2up+(\bar u-\bar v)\bar p-(\bar u+\bar v)\bar q +\bar p\bar q\over 2}\right)
\bar f(p,\bar p,\bar q)
\\
&\simeq z \int d^2\bar p d^2\bar q \exp\left\{ {1+z^{1\over2}\over 1-z^{1\over2}} y\sigma^z \bar p + \left[ 1+ \left(1+z^{1\over 2}\over 1-z^{1\over 2}\right)^2 \right] {\bar p\bar q\over 2}  \right\}
\bar f\left(-y-{1+z^{1\over2}\over 1-z^{1\over 2}}\sigma^z\bar q,\bar p, \bar q\right)
\\
&\to  z \int d^2\bar p d^2\bar q\, e^{(1+\epsilon)\bar p\bar q+2\sqrt{z}y\sigma^z \bar p }
\bar f\left(-\sigma^z\bar q,\bar p, \bar q-\sigma^z y\right)
\fe
In the last step, we again take $z\to 0$ while keeping $\sqrt{z}y$ fixed. It may seem that this limit is not well defined, because of the $y$ dependence in $\bar f\left(-\sigma^z\bar q,\bar p, \bar q-\sigma^z y\right)$.
We will see below that in fact this is not the case.

To be more precise, let us define
\ie
\lim_{z\to 0^+} z^{-1}F(z, z^{-{1\over 2}}y_\A) = \tilde F(2y_\A)
\fe
whose order ${\cal O}(y^{2s})$ term contains the boundary expectation value of the spin-$s$ component of $B$ field, with the power of $z$ stripped off. $\tilde F$ is then computed from\footnote{The limit $\xi \rightarrow \infty$ arises from the $y$-dependence in $\bar f\left(-\sigma^z\bar q,\bar p, \bar q-\sigma^z y\right)$, when taking the $z\rightarrow 0$ limit with $\sqrt{z}y$ fixed. Alternatively, one may also denote $\xi=\frac{1}{2\epsilon}$ and take a single limit $\epsilon \rightarrow 0^+$.}
\ie
& \tilde F(w_\A) =  \lim_{\epsilon\to 0^+}\left[ \int d^2y d^2z e^{(1+\epsilon) yz + zw} f(y,-\sigma^z z,z) + \lim_{\xi\to +\infty} \int d^2\bar y d^2\bar z e^{(1+\epsilon)\bar y \bar z+\bar y \sigma^z w} \bar f(-\sigma^z \bar z,\bar y,\bar z - \xi\sigma^z w) \right]
\fe
Although not obvious from this expression, the $\xi\to +\infty$ limit of the second integral is expected to be well defined, as shown below. Recall that
\ie
f(y,-\sigma^z z,z;U,V) = \int_0^1 dt\int_0^\infty d\eta (zu)
e^{(\eta u+tz)(y-v) + z \sigma^z\bar u} \sinh\left( z(tu+\sigma^z\bar v) \right).
\fe
We can compute the integrals
\ie
& \lim_{\epsilon\to 0^+}\int d^2y d^2 z e^{(1+\epsilon)yz+zw}f(y,-\sigma^z z,z;U,V)
\\
&= \lim_{\epsilon\to 0^+} \int d^2z\,(zu) \int_0^1 dt\int_0^\infty d\eta\,\delta(\eta u+(t-1-\epsilon)z)
e^{z(w-v+\sigma^z\bar u) } \sinh(z(tu+\sigma^z\bar v))
\\
&=0,
\fe
and
\ie
&\lim_{\epsilon\to 0^+} \lim_{\xi \to +\infty}\int d^2\bar y d^2\bar z e^{(1+\epsilon)\bar y\bar z+\bar y\sigma^z w}\bar f(-\sigma^z \bar z,\bar y, \bar z-\xi \sigma^z w;U,V)
\\
&= \lim_{\epsilon\to 0^+}\lim_{\xi \to +\infty} \int d^2\bar z\,((\bar z+\xi w\sigma^z)\bar u) \int_0^1 dt\int_0^\infty d\eta\,\delta(\eta \bar u+t(\bar z-\xi \sigma^z w)-((1+\epsilon)\bar z+\sigma^z w)) \\
&~~~\times e^{-((1+\epsilon)\bar z-w\sigma^z)\bar v+\bar z\sigma^z u } \sinh(t(\bar z+\xi w\sigma^z)\bar u+\bar z\sigma^z v)
\\
&=  {\rm sgn}(w\sigma^z\bar u) \int d^2\bar z\, \theta({w\sigma^z \bar z\over w\sigma^z \bar u}) \theta({(\bar z-w\sigma^z)\bar u\over w\sigma^z \bar u})
e^{\bar z(-\bar v+\sigma^z u)+ w \sigma^z \bar v } \sinh(\bar z(\bar u+\sigma^z v)-w\sigma^z\bar u).
\fe
To obtain the last line, we have used the two-dimensional $\delta$-function to integrate over $t,\eta$. The step functions come from requiring that the value of $t,\eta$ which solve the $\delta$-function constraint lie inside the corresponding integration domains.

Writing $\bar z=-(\tau_1+1) \sigma^z w + \tau_2 \bar u$, $\A_\pm = (\bar v-\sigma^z u)\pm (\bar u+\sigma^z v)$, we can express the above integral as
\ie\label{inttmp}
& -{1\over 2}(w\sigma^z \bar u)\int_0^\infty d\tau_1 \int_0^\infty d\tau_2 \left[ e^{-(\tau_1+1)(w\sigma^z\A_+)-\tau_2 (\bar u\A_+)+w\sigma^z (\bar u+\bar v) } - e^{-(\tau_1+1)(w\sigma^z\A_-)-\tau_2 (\bar u\A_-)-w\sigma^z (\bar u-\bar v) } \right]
\\
&=-{1\over 2}(w\sigma^z\bar u)\left[ {e^{-w\sigma^z(\A_+-\bar u-\bar v)}\over (w\sigma^z\A_+)(\bar u\A_+)} -{e^{-w\sigma^z(\A_-+\bar u-\bar v)}\over (w\sigma^z\A_-)(\bar u\A_-)} \right].
\fe
From (\ref{Bprime2}) and (\ref{B2phys}), we then obtain the result
\ie\label{ftd}
&\lim_{z\to 0} z^{-1}B^{(2)}(\vec x=0,z|y=z^{-{1\over 2}}\frac{w}{2},\bar y=Z=0)\\
&=-\int d^4u d^4v e^{uv-\bar u\bar v}B'^{(1)}(u,\bar u)B'^{(1)}(v,\bar v) \,(w\sigma^z\bar u)\left[ {e^{-w\sigma^z(\A_+-\bar u-\bar v)}\over (w\sigma^z\A_+)(\bar u\A_+)} - {e^{-w\sigma^z(\A_-+\bar u-\bar v)}\over (w\sigma^z\A_-)(\bar u\A_-)} \right].
\fe
The integration over $(u,\bar u,v,\bar v)$ should be understood as a contour integral, and the choice of contour is now important. The need for this choice of contour is possibly due to the slightly singular nature of the $W=0$ gauge. In the next section, we will see that the three point function is essentially
a twistor transform of
\ie\label{uvuv}
e^{uv-\bar u\bar v} (w\sigma^z\bar u)\left[ {e^{-w\sigma^z(\A_+-\bar u-\bar v)}\over (w\sigma^z\A_+)(\bar u\A_+)} -{e^{-w\sigma^z(\A_-+\bar u-\bar v)}\over (w\sigma^z\A_-)(\bar u\A_-)} \right].
\fe
Namely if we regard $(u,\bar u,v,\bar v)$ as independent holomorphic variables, and Fourier transform two of them, then we obtain (a generating function of) the three-point functions in terms of
polarization spinors (see eq. (\ref{twprop}) and the paragraph thereafter). $w=2y$ will be identified with the polarization spinor of the third (outcoming) operator.
The question of contour prescription now amounts to choosing a 4-dimensional contour (on two of $(u,\bar u,v,\bar v)$) for the twistor transform. We will demand that $w\sigma^z\A_\pm$ and $\bar u\A_\pm$ encircle the origin in the complex plane with opposite orientation, so that
\ie
{(w\sigma^z\bar u)\over (w\sigma^z\A_\pm)(\bar u\A_\pm)}
\label{rational}
\fe
picks up residue $\mp 1$ when integrated in $\A_\pm$.\footnote{One might contemplate the alternative possibility of choosing the orientation of the contour so that (\ref{rational}) picks up residue $+1$ (or $-1$) when integrated in $\A_\pm$. In this case, however, one finds that (\ref{B2-final}) vanishes identically, see eq. (\ref{solve-deltas}).}  Consequently, (\ref{ftd}) can be replaced by the residue contribution
\ie
&\lim_{z\to 0} z^{-1}B^{(2)}(\vec x=0,z|y=z^{-{1\over 2}}\frac{w}{2},\bar y=Z=0)\\
&=\int d^4u d^4v e^{uv-\bar u\bar v}B'^{(1)}(u,\bar u)B'^{(1)}(v,\bar v)\\
&~~~\times \left[ e^{-w\sigma^z(\bar u-\bar v)}\delta(\bar u-\bar v+\sigma^z(u+v))
+ e^{w\sigma^z (\bar u+\bar v)}\delta(\bar u+\bar v+\sigma^z(- u+ v)) \right].
\label{B2-final}
\fe

\section{Three point functions from twistor space}

In this section we show that a drastic simplification occurs if we consider a twistor transform of the correlation functions on the polarization spinors $\lambda_1,\lambda_2$ of the boundary sources (recall that these are related to the null polarization vectors by $({\slash\!\!\!\varepsilon}\sigma^z)_{\da\db} =\bar \lambda_\da \bar\lambda_\db$, and $\bar\lambda = \sigma^z \lambda$).

To see this, let us perform the Fourier transform of the boundary-to-bulk propagator for $B'$ in the $W=0$ gauge, with boundary source located at $\vec{x}_0$ 
\ie
B'^{(1)}(y,\bar y;\lambda) = {1\over x_0^2+1} e^{-y\left(\sigma^z+ 2 {\vec{x}_0 \cdot \vec{\sigma}-\sigma^z\over x_0^2+1}\right)\bar y}
\left\{ \exp\left[ - 2 y{\vec{x}_0 \cdot \vec{\sigma}-\sigma^z\over x_0^2+1}\bar \lambda \right]
+ \exp\left[ - 2 \bar y{\vec{x}_0 \cdot \vec{\sigma}-\sigma^z\over x_0^2+1} \lambda \right] \right\},
\label{b-t-b-generating}
\fe
whose Fourier transform is given by
\ie
&B_{tw}^{(1)}(y,\bar y;\mu) = {1\over 4} \int d^2\lambda e^{2\lambda\mu} B'^{(1)}(y,\bar y;\lambda)
\\
&=  {1\over x_0^2+1} \exp\left[ -y\left(\sigma^z+ 2 {\vec{x}_0 \cdot \vec{\sigma}-\sigma^z\over x_0^2+1}\right)\bar y \right]
\left[ \delta\left(\mu + \sigma^z{\vec{x}_0 \cdot \vec{\sigma}-\sigma^z\over x_0^2+1} y \right)
+ \delta\left(\mu - {\vec{x}_0 \cdot \vec{\sigma}-\sigma^z\over x_0^2+1} \bar y \right) \right]
\\
&= {1\over x_0^2+1}\delta\left(\mu + \sigma^z{\vec{x}_0 \cdot \vec{\sigma}-\sigma^z\over x_0^2+1} y \right)
\exp\left[ \mu\sigma^z(\vec{x}_0 \cdot \vec{\sigma}-\sigma^z)\left(\sigma^z+ 2 {\vec{x}_0 \cdot \vec{\sigma}-\sigma^z\over x_0^2+1}\right)\bar y \right] \\
&~~+ {1\over x_0^2+1} \delta\left(\mu - {\vec{x}_0 \cdot \vec{\sigma}-\sigma^z\over x_0^2+1} \bar y \right)
 \exp\left[ -y\left(\sigma^z+ 2 {\vec{x}_0 \cdot \vec{\sigma}-\sigma^z\over x_0^2+1}\right)(\vec{x}_0 \cdot \vec{\sigma}-\sigma^z)\mu \right]
 \\
 &=\delta\left( y + (\vec{x}_0 \cdot \vec{\sigma}\sigma^z-1)\mu  \right)
e^{ -\mu ( \vec{x}_0 \cdot \vec{\sigma}-\sigma^z )\bar y } + \delta\left(\bar y-(\vec{x}_0 \cdot \vec{\sigma}-\sigma^z) \mu \right)
e^{ -y(\sigma^z\vec{x}_0 \cdot \vec{\sigma}+1)\mu }.
\label{lam-to-mu}
\fe
As remarked earlier, the fact that we only have integer spins in the spectrum implies that we should actually take the contribution even in $\lambda$ in (\ref{b-t-b-generating}), or even in $\mu$ in (\ref{lam-to-mu}).
 
We can also write $\bar\mu=-\sigma^z\mu$ and
\ie
&B_{tw}^{(1)}(y,\bar y;\mu) = \delta(y- (\vec{x}_0 \cdot \vec{\sigma}-\sigma^z)\bar \mu)
e^{-\mu ( \vec{x}_0 \cdot \vec{\sigma}-\sigma^z )\bar y } + \delta\left(\bar y-(\vec{x}_0 \cdot \vec{\sigma}-\sigma^z) \mu \right)
e^{- \bar\mu (\vec{x}_0 \cdot \vec{\sigma}-\sigma^z)y }.
\fe
Let us further define
\ie
&\chi = (\vec{x}_0 \cdot \vec{\sigma}-\sigma^z)\bar\mu,\\
&\bar\chi =  (\vec{x}_0 \cdot \vec{\sigma}-\sigma^z)\mu,
\label{chi-def}
\fe
so we end up with simply
\ie\label{twprop}
&B_{tw}^{(1)}(y,\bar y;\chi,\bar\chi) = \delta(y- \chi)
e^{\bar\chi\bar y } + \delta\left(\bar y-\bar\chi \right)
e^{\chi y }.
\fe
We could regard $y,\bar y$ as independent holomorphic variables, and interpret the two terms in $B_{tw}^{(1)}$ as delta functions in the corresponding twistor space, where one of $y$ and $\bar y$ is Fourier transformed. This explains our earlier claim that the generating function of three point functions can be viewed as a twistor transform of (\ref{uvuv}) over two of $(u,\bar u,v,\bar v)$.

Assuming the choice of contour as explained in the previous section, we can now easily compute the rescaled expectation value of the outcoming higher spin fields near the boundary (more precisely, the generalized Weyl curvature of the HS fields). Denoting the position of the two boundary sources by $\vec{x}_1$ and $\vec{x}_2$,  with $\chi_{1,2}$ defined as in (\ref{chi-def}), we have
\ie
&\lim_{z\to 0} z^{-1}B^{(2)}(\vec x=0,z|z^{-{1\over 2}}y,\bar y=Z=0;\chi_1,\chi_2)\\
&
= \int d^4u d^4v e^{uv-\bar u\bar v} B'^{(1)}_{tw}(u,\bar u,\chi_1,\bar\chi_1)|_{\chi_1-{\rm even}}\, B_{tw}'^{(1)}(v,\bar v,\chi_2,\bar\chi_2)|_{\chi_2-{\rm even}}\\
&~~~\times \left[ e^{-2y\sigma^z(\bar u-\bar v)}\delta(\bar u-\bar v+\sigma^z( u+ v))
+ e^{2y\sigma^z(\bar u+\bar v)}\delta(\bar u+\bar v+\sigma^z(- u+ v)) \right] + (1\leftrightarrow 2)
\\
&= \int d^4u d^4v e^{uv-\bar u\bar v} \left[ \delta(u-\chi_1)e^{\bar\chi_1\bar u} + \delta(\bar u-\bar \chi_1) e^{\chi_1 u} \right]_{\chi_1-{\rm even}}\left[ \delta(v-\chi_2)e^{\bar\chi_2\bar v} + \delta(\bar v-\bar \chi_2) e^{\chi_2 v} \right]_{\chi_2-{\rm even}} \\
&~~~\times \left[ e^{-2y\sigma^z(\bar u-\bar v)}\delta(\bar u-\bar v+\sigma^z( u+ v))
+ e^{2y\sigma^z(\bar u+\bar v)}\delta(\bar u+\bar v+\sigma^z(- u+ v)) \right] + (1\leftrightarrow 2)
\\
&= 2 \cosh\left({\chi_1\chi_2+\bar\chi_1\bar\chi_2}\right) \left[ e^{2y(\chi_1+\chi_2)} \delta(\chi_1+\chi_2+\sigma^z(\bar\chi_1+\bar\chi_2)) + \delta(\chi_1+\chi_2+\sigma^z(\bar\chi_1+\bar\chi_2)+2y) \right]
\\
&~~~+(\chi_1\to -\chi_1)+(\chi_2\to -\chi_2)+(\chi_1\to -\chi_1,\chi_2\to -\chi_2).
\label{solve-deltas}
\fe
In terms of $\mu_1, \mu_2$, it is
\ie\label{resa}
&\lim_{z\to 0} z^{-1}B^{(2)}(\vec x=0,z|z^{-{1\over 2}}y,\bar y=Z=0;\chi_1,\chi_2)\\
&={1\over 2} \cosh\left({2\mu_1\sigma^z{\bf x_{12}}\mu_2} \right) \left[ e^{2y(\mu_1+\mu_2)} \delta({\bf x_1}\mu_1+{\bf x_2}\mu_2) + \delta({\bf x_1}\mu_1+{\bf x_2}\mu_2+\sigma^z y) \right]
\\
&~~~+(\mu_1\to -\mu_1)+(\mu_2\to -\mu_2)+(\mu_1\to -\mu_1,\mu_2\to -\mu_2).
\fe
Here and in what follows, we will use for convenience the notation ${\bf x_{1,2}}$ instead of $\vec{x}_{1,2}\cdot \vec{\sigma}$, which is equivalent since $x_{1,2}$ are by definition three dimensional vectors.

Now let us Fourier transform back in the polarization spinors $\lambda_1,\lambda_2$. For example, in the case when the outcoming field is a scalar, we can set $y=0$ in (\ref{resa}). The result is then the $(\lambda_1,\lambda_2)$-even part of
\ie
&4\int d^2\mu_1 d^2\mu_2 e^{2\mu_1\sigma^z {\bf x_{12}}\mu_2 -2\lambda_1\mu_1-2\lambda_2\mu_2}\delta({\bf x_1}\mu_1+{\bf x_2}\mu_2) + (1\leftrightarrow 2)
\\
&={4\over x_2^2}\int d^2\mu_1 \exp\left[ -{2\over x_2^2}\mu_1\sigma^z{\bf x_{12}}{\bf x_2}{\bf x_1}\mu_1
-2(\lambda_1 - x_2^{-2}\lambda_2 {\bf x_2x_1})\mu_1 \right]+ (1\leftrightarrow 2)
\\
&= {2\over  |x_1||x_2||x_{12}|} \exp\left[ {(\lambda_1 - x_2^{-2}\lambda_2 {\bf x_2x_1}) {\bf x_1 x_2 x_{12}}\sigma^z
 (\lambda_1 - x_2^{-2}{\bf x_1x_2}\lambda_2) \over 2x_1^2 x_{12}^2} \right]+ (1\leftrightarrow 2)
\\
&= {4\over |x_1||x_2| |x_{12}|} \cosh\left[ {(\hat\lambda_1 - \hat\lambda_2 )\sigma^z
(x_1^2 {\bf x_2}-x_2^2{\bf x_1}) (\hat \lambda_1 - \hat \lambda_2) \over 2x_{12}^2} \right]
\fe
where we redefined
\ie
\hat\lambda_i = {{\bf x_i}\lambda_i\over x_i^2},~~~\hat x_i = {\vec x_i\over |x_i|}.
\fe
In terms of the polarization vectors $\varepsilon_1,\varepsilon_2$, or the corresponding hatted variables, it is given by
\ie
{4\over |x_1||x_2| |x_{12}|} \cosh\left[ {{\rm Tr}
(x_1^2 {\bf x_2}-x_2^2{\bf x_1}) ({\slash\!\!\!\hat\varepsilon}_1+{\slash\!\!\!\hat\varepsilon}_2) \over 2x_{12}^2} \right] \cosh \left[{\sqrt{{\rm Tr} {\slash\!\!\!\hat\varepsilon}_1(x_1^2 {\bf x_2}-x_2^2{\bf x_1}){\slash\!\!\!\hat\varepsilon}_2(x_1^2 {\bf x_2}-x_2^2{\bf x_1}) } \over x_{12}^2}\right]
\fe

For general spin we need to keep the $y$ dependence of the outcoming field in (\ref{resa}), and hence the three point function receives two contributions, from the two terms in the second line of (\ref{resa}). The Fourier transform of the first term into $(\lambda_1,\lambda_2)$ is
\ie
&{1\over 4|x_1||x_2| |x_{12}|}
\exp\left[ {(\hat\lambda_1 - \hat\lambda_2 - y {\bf \check X_{12}} )\sigma^z
(x_1^2 {\bf x_2}-x_2^2{\bf x_1}) (\hat \lambda_1 - \hat \lambda_2 +{\bf \check X_{12}}y) \over 2x_{12}^2} \right]  + (1\leftrightarrow 2)
\\
&= {1\over 4|x_1||x_2| |x_{12}|}
\exp\left[ -{(\lambda_1 {{\bf x_1}\over x_1^2} - \lambda_2 {{\bf x_2}\over x_2^2} + y ({{\bf x_1}\over x_1^2}-{{\bf x_2}\over x_2^2}) )\sigma^z
{\bf x_1x_{12}x_2}({{\bf x_1}\over x_1^2} \lambda_1 - {{\bf x_2}\over x_2^2} \lambda_2 +
({{\bf x_1}\over x_1^2}-{{\bf x_2}\over x_2^2})y) \over 2x_{12}^2} \right] + (1\leftrightarrow 2)
\\
&= {1\over 4|x_1||x_2| |x_{12}|}
\exp\left[ {x_2^2\lambda_1\sigma^z {\bf x_{12}x_2x_1}\lambda_1
+x_1^2\lambda_2\sigma^z{\bf x_2 x_1 x_{12}}\lambda_2 + x_{12}^2y\sigma^z{\bf x_1x_{12}x_2}y
\over 2x_1^2x_2^2x_{12}^2} \right.\\
&\left.~~~+ \lambda_1 \sigma^z{{\bf x_{12}}\over x_{12}^2}\lambda_2 + \lambda_1\sigma^z {{\bf x_1}\over x_1^2}y + \lambda_2\sigma^z {{\bf x_2}\over x_2^2}y \right] + (1\leftrightarrow 2)
\fe
where we defined ${\bf \check X_{12}} = {{\bf x_1}\over x_1^2}-{{\bf x_2}\over x_2^2}$.
Now we replace $y$ by $\lambda_3$, and replace the origin by $\vec x_3$ where the third operator is inserted. The resulting contribution to the (generating function of) three point functions is
\ie
&{1\over 4|x_{12}||x_{23}| |x_{31}|}
\exp\left[ {x_{23}^2\lambda_1\sigma^z {\bf x_{12}x_{23}x_{13}}\lambda_1
+x_{13}^2\lambda_2\sigma^z{\bf x_{23} x_{31} x_{21}}\lambda_2 + x_{12}^2\lambda_3\sigma^z{\bf x_{31}x_{12}x_{32}}\lambda_3
\over 2x_{12}^2x_{23}^2x_{31}^2} \right.\\
&\left.~~~+ \lambda_1 \sigma^z{{\bf x_{12}}\over x_{12}^2}\lambda_2 + \lambda_1\sigma^z {{\bf x_{13}}\over x_{13}^2}\lambda_3 + \lambda_2\sigma^z {{\bf x_{23}}\over x_{23}^2}\lambda_3 \right] + (1\leftrightarrow 2).
\fe
On the other hand, the Fourier transform of the second term in the second line of (\ref{resa}) is given by
\ie
&{1\over 4|x_1||x_2| |x_{12}|}
\exp\left[ {(\hat\lambda_1 - \hat\lambda_2 - y {\bf \check X_{12}} )\sigma^z
(x_1^2 {\bf x_2}-x_2^2{\bf x_1}) (\hat \lambda_1 - \hat \lambda_2 +{\bf \check X_{12}}y) \over 2x_{12}^2}
-2\lambda_2\sigma^z{{\bf x_2}\over x_2^2}y \right]  + (1\leftrightarrow 2)
\\
&\to {1\over 4|x_{12}||x_{23}| |x_{31}|}
\exp\left[ {x_{23}^2\lambda_1\sigma^z {\bf x_{12}x_{23}x_{13}}\lambda_1
+x_{13}^2\lambda_2\sigma^z{\bf x_{23} x_{31} x_{21}}\lambda_2 + x_{12}^2\lambda_3\sigma^z{\bf x_{31}x_{12}x_{32}}\lambda_3
\over 2x_{12}^2x_{23}^2x_{31}^2} \right.\\
&\left.~~~+ \lambda_1 \sigma^z{{\bf x_{12}}\over x_{12}^2}\lambda_2 + \lambda_1\sigma^z {{\bf x_{13}}\over x_{13}^2}\lambda_3 - \lambda_2\sigma^z {{\bf x_{23}}\over x_{23}^2}\lambda_3 \right] + (1\leftrightarrow 2).
\fe
where we have again made the substitution of $y$ by $\lambda_3$ so that the crossing symmetry in the three currents is manifest. Together with the terms related by flipping the sign of $\lambda_1$ and $\lambda_2$ respectively, the total contribution to the generating function of all three point functions is
\begin{eqnarray}
\label{ttm}
&&{4\over |x_{12}||x_{23}| |x_{31}|}
\cosh\left( {x_{23}^2\lambda_1\sigma^z {\bf x_{12}x_{23}x_{13}}\lambda_1
+x_{13}^2\lambda_2\sigma^z{\bf x_{23} x_{31} x_{21}}\lambda_2 + x_{12}^2\lambda_3\sigma^z{\bf x_{31}x_{12}x_{32}}\lambda_3
\over 2x_{12}^2x_{23}^2x_{31}^2} \right)\cr
&&~~~\times \cosh\left(\lambda_1 \sigma^z{{\bf x_{12}}\over x_{12}^2}\lambda_2\right)\cosh\left( \lambda_1\sigma^z {{\bf x_{13}}\over x_{13}^2}\lambda_3 \right)\cosh\left( \lambda_2\sigma^z {{\bf x_{23}}\over x_{23}^2}\lambda_3 \right).
\end{eqnarray}
A given three point function of higher spin currents $\langle J_{s_1}(x_1;\lambda_1) J_{s_2}(x_2;\lambda_2) J_{s_3}(x_3;\lambda_3)\rangle$ can be now obtained from this generating function by simply extracting the contribution which goes like $\lambda_1^{2s_1} \lambda_2^{2s_2} \lambda_3^{2s_3}$.

In the conjectured dual free scalar theory,
using free field Wick contractions, one may derive the following generating function of $n$-point functions \cite{Giombi:2009wh} (here we assume null polarization vectors as above)
\ie
\frac{1}{n} \sum_{\sigma\in S_n}
P_\sigma \overrightarrow{\prod_{i=1}^n} \left[\cos(\sqrt{4({\varepsilon}_i\cdot\overleftarrow\partial_i)
({\varepsilon}_i\cdot\overrightarrow\partial_i)})~{1\over |x_{i}
-x_{i+1}+{\varepsilon}_{i} + {\varepsilon}_{i+1}|}\right]
\label{free-gen}
\fe
where $P_\sigma$ stands for the permutation on $(\vec x_i;{\vec \varepsilon}_i)$ by $\sigma$, and the product is understood to be of cyclic order ($\overleftarrow\partial$ and $\overrightarrow\partial$ act on their neighboring propagators only). The $n$-point function for given spins is obtained by extracting the appropriate powers of the polarization vectors $\varepsilon_i$. Our bulk result (\ref{ttm}) in fact generates exactly the same set of three point functions as the $n=3$ case of (\ref{free-gen}).\footnote{Note however that the two generating functions are defined with different normalizations on the currents.} A proof is given in the appendix.
Thus we have found complete agreement of the bulk tree-level three-point functions with the three point functions of higher spin currents in the free $O(N)$ scalar CFT.
In the following we describe some simple checks in special cases.

Without loss of generality, we can fix the positions $x_1, x_2, x_3$ by conformal symmetry to $x_1=e_1,x_2=-e_1,x_3=0$, so that (\ref{ttm}) reduces to
\ie
&2\cosh\left( {-\lambda_1\sigma^z {\bf e_1}\lambda_1
-\lambda_2\sigma^z{\bf  e_1}\lambda_2 -4 \lambda_3\sigma^z{\bf e_1}\lambda_3
\over 4}\right) \cosh\left({\lambda_1 \sigma^z{{\bf e_1}}\lambda_2\over 2}\right)\cosh\left( \lambda_1\sigma^z {{\bf e_1}}\lambda_3 \right)\cosh\left( \lambda_2\sigma^z {{\bf e_1}}\lambda_3 \right) 
\fe
As an example, let us extract the three point function of the stress energy tensor $\langle TTT\rangle$,
from the ${\cal O}(\lambda_1^4\lambda_2^4\lambda_3^4)$ term. If we further use the remaining 1 conformal transformation to set $e_1\cdot \varepsilon_1=0$, we end up with the following simple expression for $\langle TTT\rangle$,
\ie\label{tttsp}
&{1\over 24} \left[  (e_1\cdot \varepsilon_2)^2 (\varepsilon_1\cdot\varepsilon_3)^2
+ (e_1\cdot \varepsilon_3)^2 (\varepsilon_1\cdot\varepsilon_2)^2
+36 (e_1\cdot \varepsilon_2)(e_1\cdot \varepsilon_3)(\epsilon_1\cdot\epsilon_2)(\epsilon_1\cdot\epsilon_3) \right.\\
&\left.- 12 (\epsilon_1\cdot\epsilon_2)(\epsilon_1\cdot\epsilon_3)(\epsilon_2\cdot\epsilon_3) \right]
\fe
Let us compare this with the stress energy tensor of a free massless scalar in 3d, contracted with a null polarization vector $\varepsilon$,
\ie
T_{\varepsilon} = (\varepsilon\cdot \partial\phi)^2 - {1\over 8}(\varepsilon\cdot\partial)^2\phi^2.
\fe
We have
\ie
&\langle T_{\varepsilon_1}(x_1)T_{\varepsilon_2}(x_2)T_{\varepsilon_3}(x_3) \rangle
= \langle (\varepsilon_1\cdot \partial\phi(x_1))^2(\varepsilon_2\cdot \partial\phi(x_2))^2(\varepsilon_3\cdot \partial\phi(x_3))^2 \rangle\\
&~~-{1\over 8}\left[ (\varepsilon_1\cdot \partial_1)^2\left\langle \phi(x_1)^2 (\varepsilon_2\cdot \partial\phi(x_2))^2(\varepsilon_3\cdot \partial\phi(x_3))^2\right\rangle + cyclic \right]
\\
&~~+{1\over 64}\left[ (\varepsilon_1\cdot \partial_1)^2(\varepsilon_2\cdot \partial_2)^2 \left\langle \phi(x_1)^2\phi(x_2)^2(\varepsilon_3\cdot \partial\phi(x_3))^2\right\rangle + cyclic \right]
\\
&~~-{1\over 512} (\varepsilon_1\cdot \partial_1)^2(\varepsilon_2\cdot \partial_2)^2(\varepsilon_3\cdot \partial_3)^2\left\langle \phi(x_1)^2 \phi(x_2)^2 \phi(x_3)^2 \right\rangle
\\
&= 8 \left(\varepsilon_1\cdot \partial_1 \varepsilon_2\cdot \partial_2 {1\over |x_{12}|}\right)
\left(\varepsilon_1\cdot \partial_1 \varepsilon_3\cdot \partial_3 {1\over |x_{13}|}\right)
\left(\varepsilon_3\cdot \partial_3 \varepsilon_2\cdot \partial_2 {1\over |x_{23}|}\right)\\
&~~-\left\{ (\varepsilon_1\cdot \partial_1)^2 \left[ \left(\varepsilon_2\cdot \partial_2{1\over |x_{12}|}\right)
\left(\varepsilon_3\cdot \partial_3{1\over |x_{13}|}\right)\left(\varepsilon_3\cdot \partial_3 \varepsilon_2\cdot \partial_2 {1\over |x_{23}|}\right)\right]
 + cyclic \right\}
\\
&~~+{1\over 8}\left\{ (\varepsilon_1\cdot \partial_1)^2(\varepsilon_2\cdot \partial_2)^2
\left[ {1\over |x_{12}|}
\left(\varepsilon_3\cdot \partial_3{1\over |x_{13}|}\right)\left(\varepsilon_3\cdot \partial_3 {1\over |x_{23}|}\right)\right] + cyclic \right\}
\\
&~~-{1\over 64} (\varepsilon_1\cdot \partial_1)^2(\varepsilon_2\cdot \partial_2)^2(\varepsilon_3\cdot \partial_3)^2 {1\over |x_{12}||x_{13}||x_{23}|}
\fe
Of course, we could also extract this result directly from the generating function (\ref{free-gen}), but we have repeated the derivation for clarity.
Without loss of generality, we can now specialize to the case $x_1=e_1,x_2=-e_1,x_3=0$ and $e_1\cdot \varepsilon_1=0$ using conformal symmetry, and the result exactly matches (\ref{tttsp}) (up to the overall normalization constant).

Another check of (\ref{ttm}) is in the limit $\vec x_{12}=\vec\delta\to 0$, $\vec x_{13}\simeq \vec x_{23}\simeq \vec x$. This can be compared to the limit of ``colliding sources" which was studied in \cite{Giombi:2009wh}. We have
\ie
{1\over x^2\delta} \exp\left( {\lambda_1\sigma^z{\slash\!\!\!\delta}\lambda_1+\lambda_2\sigma^z{\slash\!\!\!\delta}\lambda_2\over 2\delta^2} + {\lambda_3\sigma^z{\bf x}{\slash\!\!\!\delta}{\bf x}\lambda_3\over 2x^4} \right) \cosh\left[ {\lambda_1\sigma^z{\slash\!\!\!\delta}\lambda_2\over\delta^2}
+{(\lambda_1+\lambda_2)\sigma^z{\bf x}\lambda_3\over x^2} \right]\,.
\fe
There are two special cases that we studied before in the ``physical gauge": $\lambda_2=0$ and $\lambda_3=0$. In the $\lambda_2=0$ case, the three point function in the $\delta\to 0$ limit is
\ie
{1\over x^2\delta} \exp\left( {\lambda_1\sigma^z{\slash\!\!\!\delta}\lambda_1\over 2\delta^2} + {\lambda_3\sigma^z{\bf x}{\slash\!\!\!\delta}{\bf x}\lambda_3\over 2x^4} \right)
\label{collide1}
\fe
whereas in the $\lambda_3=0$ case, it is given by
\ie
&{1\over x^2\delta} \exp\left( {\lambda_1\sigma^z{\slash\!\!\!\delta}\lambda_1+\lambda_2\sigma^z{\slash\!\!\!\delta}\lambda_2\over 2\delta^2}  \right) \cosh\left( {\lambda_1\sigma^z{\slash\!\!\!\delta}\lambda_2\over\delta^2} \right)
\\
&={1\over x^2\delta} \exp\left[ {(\lambda_1+\lambda_2)\sigma^z{\slash\!\!\!\delta}(\lambda_1+\lambda_2)  \over 2\delta^2}\right].
\label{collide2}
\fe
These indeed agree with the results we found in \cite{Giombi:2009wh}. \footnote{To see this, compare (\ref{collide1}) and (\ref{collide2}) to respectively eq. (6.23) and eq. (4.88) of \cite{Giombi:2009wh}.}

\bigskip

Finally, let us turn to the type B model of \cite{Sezgin:2003pt}. Instead of (\ref{twprop}), the boundary-to-bulk propagator for the $B$ master field in the type B model, after the Fourier transform in polarization spinors, is given by\footnote{In the type B model, the third equation of (\ref{cleom}) is modified to $d_Z S+S*S=B*(-iK dz^2+i\bar K d\bar z^2)$. This leads to the extra factors of $i$ and $-i$ in the boundary-to-bulk propagator for $B$.}
\ie\label{twb}
&B_{tw;B}^{(1)}(y,\bar y;\chi,\bar\chi) = i\delta(y- \chi)
e^{\bar\chi\bar y } - i\delta\left(\bar y-\bar\chi \right)
e^{\chi y }.
\fe
Note that the scalar field component has disappeared from (\ref{twb}). The bulk scalar is parity odd in the type B model, and the ``standard" boundary condition assigns scaling dimension 2 to its dual operator. Therefore the scalar has to be treated separately, and we will only consider HS currents for now. The generating function for $\langle JJJ\rangle$ is now the $(\lambda_1,\lambda_2,\lambda_3)$-even part of
\ie\label{ttmf}
&{4\over |x_{12}||x_{23}| |x_{31}|}
\sinh\left( {x_{23}^2\lambda_1\sigma^z {\bf x_{12}x_{23}x_{13}}\lambda_1
+x_{13}^2\lambda_2\sigma^z{\bf x_{23} x_{31} x_{21}}\lambda_2 + x_{12}^2\lambda_3\sigma^z{\bf x_{31}x_{12}x_{32}}\lambda_3
\over 2x_{12}^2x_{23}^2x_{31}^2} \right)\\
&~~~\times \sinh\left(\lambda_1 \sigma^z{{\bf x_{12}}\over x_{12}^2}\lambda_2\right) \sinh\left( \lambda_1\sigma^z {{\bf x_{13}}\over x_{13}^2}\lambda_3 \right) \sinh\left( \lambda_2\sigma^z {{\bf x_{23}}\over x_{23}^2}\lambda_3 \right).
\fe
This is conjectured to be dual to the free $O(N)$ fermion theory in three dimensions \cite{Sezgin:2003pt}. As a check let us consider $\langle TTT\rangle$. As before, by conformal symmetry we can fix $x_1=e_1,x_2=-e_1,x_3=0$ and $e_1\cdot\varepsilon_1=0$, and the three point function of the stress energy tensor from Vasiliev theory in this case is given by
\ie\label{testa}
-{1\over 3}\left( e_1\cdot \varepsilon_3\varepsilon_1\cdot\varepsilon_2+e_1\cdot \varepsilon_2\varepsilon_1\cdot\varepsilon_3 \right)^2
\fe
The stress energy tensor of the free fermion theory, with null polarization vector $\varepsilon$, is
\ie
T_\varepsilon^F = \psi {\slash\!\!\!\varepsilon} (\vec\varepsilon\cdot \vec{\partial})\psi.
\fe
It is straightforward to check that that (\ref{testa}) indeed produces exactly $\langle T^F_{\varepsilon_1}(x_1)T^F_{\varepsilon_2}(x_2)T^F_{\varepsilon_3}(x_3)\rangle$, up to the overall normalization constant.

\section{Concluding remarks}

In this paper we have shown that the tree level three point functions of Vasiliev's minimal bosonic higher spin gauge theory in $AdS_4$ exactly agree with the three point functions of higher spin currents in the free theory of $N$ massless scalars in the $O(N)$ singlet sector in 3 dimensions. The bulk computation is made possible by the remarkable simplification in the $W=0$ gauge, where the integration over the $AdS_4$ is replaced by the $*$-product of twistor-like internal variables of Vasiliev's master fields.

The agreement of the three point functions $\langle JJJ\rangle$ with the complete position and polarization dependence included is a nontrivial check of the conjecture of Sezgin-Sundell-Klebanov-Polyakov. As a special case, the three point function of the stress energy tensor $\langle TTT\rangle$ in a three dimensional CFT is constrained by conformal symmetry up to a linear combination of two possible structures, one corresponding to that of a free massless scalar, the other corresponding to that of a free massless fermion \cite{Osborn:1993cr}. From the perspective of the bulk Lagrangian, the tree level $\langle TTT\rangle$ is sensitive to the higher derivative terms in the graviton. Indeed, computing $\langle TTT\rangle$ from pure Einstein gravity in $AdS_4$ would produce a linear combination of the two tensor structures \cite{Arutyunov:1999nw}. The agreement we found is therefore a test of the precise higher derivative structure of Vasiliev's theory.

We have also seen that the three point functions in type B model matches that of free fermions, verifying a conjecture of \cite{Sezgin:2003pt}. In fact, our result also applies to the nonminimal Vasiliev theory, without imposing the projection (\ref{proja}) and so both even and odd integer spins are included. The result then matches the free CFT of $N$ complex scalars in the $SU(N)$ singlet sector. In this theory, we may choose alternative boundary conditions for the bulk scalar field as well as the vector gauge field \cite{Witten:2003ya, Leigh:2003ez, Petkou:2004wb}, which would lead to conjectured dual critical scalar QED (with $N$ flavors) or critical $\mathbb{CP}^{N-1}$ models in 2+1 dimensions. It would be very interesting if one can learn about these CFTs from Vasiliev theory.

It is now technically feasible to generalize our computation to higher point functions as well as to loop corrections in the bulk. An extremely interesting problem is to understand the HS symmetry breaking in the critical $O(N)$ model from corrections by scalar loops with $\Delta=2$ boundary condition in the bulk. We hope to report on results toward these directions in the near future.

\subsection*{Acknowledgments}

We are grateful to N. Boulanger, V.E. Didenko, C. Iazeolla, I. Klebanov, J. Maldacena for very useful discussions, and especially to P. Sundell for pointing out to us the importance of the $W=0$ gauge in Vasiliev theory.
X.Y. would like to thank Universit\'e de Mons and University of Crete for their hospitality during the course of this work.
This work is supported in part by the Fundamental Laws Initiative Fund at Harvard University. S.G. is supported
in part by NSF Award DMS-0244464. X.Y. is supported in part by NSF Award PHY-0847457.

\appendix

\section{The equivalence of two generating functions}

In this appendix we will show that (\ref{ttm}) and (\ref{free-gen}) generate the same three-point functions of higher spin currents. In terms of the null polarization vectors $\vec\varepsilon_i$, (\ref{ttm}) can be written as
\ie
&{4\over |x_{12}||x_{23}||x_{31}|} \cosh \left[ \left( {\vec x_{13}\over x_{13}^2}-{\vec x_{12}\over x_{12}^2} \right)\cdot \vec\varepsilon_1 + \left( {\vec x_{21}\over x_{21}^2}-{\vec x_{23}\over x_{23}^2} \right)\cdot \vec\varepsilon_2 + \left( {\vec x_{32}\over x_{32}^2}-{\vec x_{31}\over x_{31}^2} \right)\cdot \vec\varepsilon_3 \right]\\
&\times \prod_{i=1}^3 \cosh\left[ {2\over x_{i,i+1}^2}\sqrt{{(\varepsilon_i\cdot \vec x_{i,i+1})(\varepsilon_{i+1}\cdot \vec x_{i,i+1})} - {1\over 2}x_{i,i+1}^2{\vec\varepsilon_i\cdot\vec\varepsilon_{i+1}} } \right].
\fe
We can use the conformal group to fix $\vec\varepsilon_i=t_i \vec\varepsilon$, $i=1,2,3$, where $t_i$ is a scale factor and $\vec\varepsilon$ is a common polarization vector. The expression then simplifies to
\ie\label{simp}
{1\over 2|x_{12}||x_{23}||x_{31}|} \sum_{\eta_i=\pm 1}\cosh\left[ {\vec\varepsilon\cdot\vec x_{12}\over x_{12}^2}({t_1}^{1\over 2}+\eta_3 {t_2}^{1\over 2})^2 + {\vec\varepsilon\cdot\vec x_{23}\over x_{23}^2}({t_2}^{1\over 2}+\eta_1{t_3}^{1\over 2})^2 + {\vec\varepsilon\cdot\vec x_{31}\over x_{31}^2}({t_3}^{1\over 2}+\eta_2{t_1}^{1\over 2})^2 \right]
\fe
On the other hand, (\ref{free-gen}) for $n=3$ can be written as
\ie
&{1\over 8}\sum_{\eta_i=\pm 1} \overrightarrow{\prod_{i=1}^3}  \exp\left[\left(\sqrt{\varepsilon_i\cdot\overrightarrow\partial_i}+\eta_i \sqrt{-\varepsilon_i\cdot\overleftarrow\partial_i}\right)^2\right]~{1\over |x_{i,i+1}|} + (1\leftrightarrow 2)
\\
&\to {1\over 8}\sum_{\eta_i=\pm 1} \overrightarrow{\prod_{i=1}^3}  \exp\left[t_i\left(\sqrt{\varepsilon\cdot\overrightarrow\partial_i}+\eta_i \sqrt{-\varepsilon\cdot\overleftarrow\partial_i}\right)^2\right]~{1\over |x_{i,i+1}|}+ (1\leftrightarrow 2)
\fe
where in the second step we have restricted to the case $\vec\varepsilon_i=t_i\vec\varepsilon$. Expanding the exponential, we have
\ie
&{1\over 8}\sum_{\eta_i=\pm1}\sum_{s_1,s_2,s_3}{t_1^{s_1}t_2^{s_2}t_3^{s_3}\over s_1!s_2!s_3!} \sum_{n_1,n_2,n_3} {2s_1\choose n_1}{2s_2\choose n_2}{2s_3\choose n_3} \eta_1^{n_1}\eta_2^{n_2}\eta_3^{n_3}\left[(\varepsilon\cdot\vec\partial_1)^{s_1-{n_1\over 2}+{n_2\over 2}}{1\over |x_{12}|}\right] \\
&~~\times  \left[(\varepsilon\cdot\vec\partial_2)^{s_2-{n_2\over 2}+{n_3\over 2}}{1\over |x_{23}|}\right] \left[(\varepsilon\cdot\vec\partial_1)^{s_3-{n_3\over 2}+{n_1\over 2}}{1\over |x_{31}|}\right] + (1\leftrightarrow 2)
\\
&= {1\over |x_{12}||x_{23}||x_{31}|} \sum_{s_1,s_2,s_3}{t_1^{s_1}t_2^{s_2}t_3^{s_3}\over s_1!s_2!s_3!} \sum_{m_1,m_2,m_3} {2s_1\choose 2m_1}{2s_2\choose 2m_2}{2s_3\choose 2m_3} 2^{s_1+s_2+s_3} \\
&~~\times  \left[ {\Gamma(s_1-m_1+m_2+{1\over 2})\over\Gamma({1\over 2})} ({\varepsilon\cdot\vec x_{12}\over x_{12}^2})^{s_1-m_1+m_2} \right]\left[ {\Gamma(s_2-m_2+m_3+{1\over 2})\over\Gamma({1\over 2})}  ({\varepsilon\cdot\vec x_{23}\over x_{23}^2})^{s_2-m_2+m_3} \right]\\
&~~\times \left[{\Gamma(s_3-m_3+m_1+{1\over 2})\over\Gamma({1\over 2})} ({\varepsilon\cdot\vec x_{31}\over x_{31}^2})^{s_3-m_3+m_1} \right]+ (1\leftrightarrow 2)
\fe
Redefining $s_1-m_1+m_2=k_1$, $s_2-m_2+m_3=k_2$, $s_3-m_3+m_1=k_3$, we can write it as
\ie
& {1\over |x_{12}||x_{23}||x_{31}|} \sum_{s_1,s_2,s_3}{t_1^{s_1}t_2^{s_2}t_3^{s_3}\over s_1!s_2!s_3!} \sum_{m_1,m_2,m_3} {2s_1\choose 2m_1}{2s_2\choose 2m_2}{2s_3\choose 2m_3} 2^{k_1+k_2+k_3} \\
&~~\times  \left[ {\Gamma(k_1+{1\over 2})\over\Gamma({1\over 2})} ({\varepsilon\cdot\vec x_{12}\over x_{12}^2})^{k_1} \right]\left[ {\Gamma(k_2+{1\over 2})\over\Gamma({1\over 2})}  ({\varepsilon\cdot\vec x_{23}\over x_{23}^2})^{k_2} \right] \left[{\Gamma(k_3+{1\over 2})\over\Gamma({1\over 2})} ({\varepsilon\cdot\vec x_{31}\over x_{31}^2})^{k_3} \right] + (1\leftrightarrow 2)
\\
& = \sum_{s_1,s_2,s_3}{t_1^{s_1}t_2^{s_2}t_3^{s_3}} A_{s_1,s_2,s_3}(\vec x_i, \varepsilon)
\fe
The term $A_{s_1,s_2,s_3}(\vec x_i,\varepsilon)$ gives the three-point function of currents of spin $(s_1,s_2,s_3)$, $\langle J_{s_1} J_{s_2} J_{s_3} \rangle$, up to a normalization factor. Now consider a sum with a different normalization factor on the currents,
\ie
&\sum_{s_1,s_2,s_3} {2^{s_1+s_2+s_3}s_1!s_2!s_3!\over (2s_1)!(2s_2)!(2s_3)!}{t_1^{s_1}t_2^{s_2}t_3^{s_3}} A_{s_1,s_2,s_3}(\vec x_i, \varepsilon)
\\
&= {1\over 8|x_{12}||x_{23}||x_{31}|}\sum_{\eta_i=\pm1} \sum_{k_1,k_2,k_3} (t_1^{1\over 2}+\eta_3 t_2^{1\over 2})^{2k_1} (t_2^{1\over 2}+\eta_1 t_3^{1\over 2})^{2k_2} (t_3^{1\over 2}+\eta_2 t_1^{1\over 2})^{2k_3}
4^{k_1+k_2+k_3} \\
&~~\times  \left[ {\Gamma(k_1+{1\over 2})\over\Gamma({1\over 2}) (2k_1)!} ({\varepsilon\cdot\vec x_{12}\over x_{12}^2})^{k_1} \right]\left[ {\Gamma(k_2+{1\over 2})\over\Gamma({1\over 2}) (2k_2)!}  ({\varepsilon\cdot\vec x_{23}\over x_{23}^2})^{k_2} \right] \left[{\Gamma(k_3+{1\over 2})\over\Gamma({1\over 2})(2k_3)!} ({\varepsilon\cdot\vec x_{31}\over x_{31}^2})^{k_3} \right]+ (1\leftrightarrow 2)
\\
&= {1\over 8|x_{12}||x_{23}||x_{31}|}\sum_{\eta_i=\pm 1}\exp\left[ {\vec\varepsilon\cdot\vec x_{12}\over x_{12}^2}({t_1}^{1\over 2}+\eta_3 {t_2}^{1\over 2})^2 + {\vec\varepsilon\cdot\vec x_{23}\over x_{23}^2}({t_2}^{1\over 2}+\eta_1{t_3}^{1\over 2})^2 + {\vec\varepsilon\cdot\vec x_{31}\over x_{31}^2}({t_3}^{1\over 2}+\eta_2{t_1}^{1\over 2})^2 \right]\\
&~~+ (1\leftrightarrow 2)
\\
&= {1\over 4|x_{12}||x_{23}||x_{31}|}\sum_{\eta_i=\pm 1}\cosh\left[ {\vec\varepsilon\cdot\vec x_{12}\over x_{12}^2}({t_1}^{1\over 2}+\eta_3 {t_2}^{1\over 2})^2 + {\vec\varepsilon\cdot\vec x_{23}\over x_{23}^2}({t_2}^{1\over 2}+\eta_1{t_3}^{1\over 2})^2 + {\vec\varepsilon\cdot\vec x_{31}\over x_{31}^2}({t_3}^{1\over 2}+\eta_2{t_1}^{1\over 2})^2 \right]
\fe
This indeed agrees with (\ref{simp}), thus proving the equivalence of the generating functions.

\end{document}